\documentclass[english,aps,preprint]{revtex4}
\usepackage{amsmath}
\usepackage{amssymb}

\makeatletter

\providecommand{\tabularnewline}{\\}

\usepackage{babel}
\makeatother
\begin{document}

\preprint{BROWN-HET-1445, ICN-UNAM-05/05A}

\title{A prototype for dS/CFT}

\author{Alberto G\"uijosa}

\email{alberto@nucleares.unam.mx}

\affiliation{\textsl{Departamento de F\'{\i}sica de Altas Energ\'{\i}as, Instituto
de Ciencias Nucleares, Universidad Nacional Aut\'onoma de M\'exico,
Apdo. Postal 70-543, M\'exico D.F. 04510, MEXICO}}

\author{David A. Lowe}

\author{Jeff Murugan}

\email{lowe@brown.edu, jeff@het.brown.edu}

\affiliation{\textsl{Department of Physics,  Brown University, Providence, RI
02912, USA}}

\begin{abstract}
We consider  $dS_{2}/CFT_{1}$ where the asymptotic symmetry group
of the de Sitter spacetime contains the Virasoro algebra. We construct
representations of the Virasoro algebra realized in the Fock space
of a massive scalar field in de Sitter, built as excitations of the
Euclidean vacuum state. These representations are unitary, without
highest weight, and have vanishing central charge. They provide a
prototype for a new class of conformal field theories dual to de Sitter
backgrounds in string theory. The mapping of operators in the CFT
to bulk quantities is described in detail. We comment on the extension
to $dS_{3}/CFT_{2}$.
\end{abstract}
\maketitle

\section{introduction}

The main tool string theorists have at their disposal in understanding
nonperturbative spacetime physics is duality with large $N$ quantum
field theories, such as AdS/CFT and Matrix theory. The generalization
of these methods to de Sitter space \cite{Strominger:2001pn} confronts
us with various interesting challenges. In particular, the modes of
a generic massive scalar field in $d$-dimensional de Sitter transform
under the isometry group $SO(d,1)$ as principal series representations
\cite{Chernikov:1968zm,Tagirov:1972vv}, so the corresponding operators
in the dual CFT must transform in the same way under the global conformal
group \cite{Guijosa:2003ze,Lowe:2004nw,Balasubramanian:2002zh,Leblond:2002tf}.
Such representations do not appear in the conformal field theories
common in string theory and statistical mechanics. It is therefore
an important open problem to better understand this new class of conformal
field theories.

In this paper, we construct perhaps the simplest example of a free
conformal field theory that gives rise to principal series representations.
We begin with two-dimensional de Sitter where the $SL(2,\mathbb{R})$
isometry group appears as a subgroup of a much larger asymptotic symmetry
algebra, the Virasoro algebra. The representations of the full Virasoro
algebra are realized on the multiparticle Fock space of a free massive
scalar field in de Sitter %
\footnote{Related work by Neretin extends the complementary series representations
of $SL(2,\mathbb{R})$ to the Virasoro algebra with vanishing central
charge \cite{Neretin1982,Neretin1983,Neretin1994}.%
}, and are consequently unitary. The Virasoro algebra is found to close
with vanishing central term. The representations are presented in
a basis obtained in a spherical slicing of de Sitter, and in another
basis obtained using a flat slicing. 

We then consider the analogous computation for the case of three-dimensional
de Sitter with a flat slicing, where the $SL(2,\mathbb{C})$ isometry
group lifts to an asymptotic symmetry group containing commuting left/right
copies of the Virasoro algebra. The Virasoro currents are well-defined
locally, but we find only the global $SL(2,\mathbb{C})$ subgroup
gives rise to properly-defined charges. The bulk matter modes therefore
do not yield representations of Virasoro. The local Virasoro symmetry
nevertheless constrains the properties of CFT correlators, as in standard
CFT.

We briefly consider the problem of counting the horizon entropy in
de Sitter using this framework. On a positive note, the one-to-one
mapping between CFT operators and scalar particle states allows one
to explicitly map the thermal entropy of a static patch observer into
a CFT computation. Other authors have speculated that applying the
Cardy formula to this counting problem will yield the correct, finite
horizon entropy of de Sitter \cite{Park:1998qk,Balasubramanian:2001nb,Bousso:2001mw,Park:2003hr}.
However since the mapping between the CFT counting problem and the
bulk scalar field counting problem really is one-to-one, the result
is divergent. We speculate on further physical input needed to regulate
this result.

\section{Basic setup}

Let us collect together a few elementary results on mode expansions
of a scalar field in $d$-dimensional de Sitter space (further details
may be found in \cite{Bousso:2001mw}). Throughout the paper we will
take the mass of the field to lie in the generic range $m>(d-1)/2$,
which is the range associated with principal unitary series representations.
Moreover, we will assume all expansions are made around the Bunch-Davies/Euclidean
vacuum state.

\subsection{Spherical slicing}

The spacetime metric is \[
ds^{2}=-d\tau^{2}+\cosh^{2}\tau~d\Omega_{d-1}^{2}~,\]
with $d\Omega_{d-1}^{2}$ the standard metric on the $d-1$ sphere.
These coordinates cover the complete de Sitter spacetime. 

A massive scalar field has a mode expansion in terms of creation/annihilation
operators $a_{lj}$ \begin{equation}
\phi=\sum_{l,j}y_{l}(\tau)Y_{lj}(\Omega)a_{lj}+\mathrm{h.c.}~,\label{eq:modede}\end{equation}
where the $Y_{lj}(\Omega)$ are spherical harmonics, labeled by a
non-negative integer $l$ and an index $j=(j_{1},\cdots,j_{d-2})$.
The $a_{lj}$ annihilate the Euclidean vacuum. The $y_{l}$'s are
solutions to a hypergeometric equation. We will only need their expansion
in the limit $\tau\to-\infty$\begin{equation}
y_{l}(\tau)\sim A_{l}e^{h_{-}\tau}+B_{l}e^{h_{+}\tau},\label{eq:sphmod}\end{equation}
with \begin{equation}
h_{\pm}=\frac{d-1}{2}\pm i\mu,\qquad\mu=\sqrt{{m^{2}-\frac{(d-1)^{2}}{4}}}~,\label{eq:hpmdef}\end{equation}
and $A_{l}$ and $B_{l}$ constants\[
A_{l}=\frac{2^{\frac{d-2}{2}}e^{i\theta_{l}}}{\sqrt{\mu(1-e^{-2\pi\mu})}}~,\qquad B_{l}=\frac{2^{\frac{d-2}{2}}e^{-i\theta_{l}+i\pi(L+\frac{d-1}{2})}}{\sqrt{\mu(e^{2\pi\mu}-1)}}~,\]
defined in terms of the phase\[
e^{-2i\theta_{l}}=e^{i\pi(l-\frac{d-1}{2})}\frac{\Gamma(-i\mu)\Gamma(l+\frac{d-1}{2}+i\mu)}{\Gamma(i\mu)\Gamma(l+\frac{d-1}{2}-i\mu)}~.\]
The normalization is chosen so that the modes are orthonormal with
respect to the Klein-Gordon inner product\begin{equation}
(\psi,\phi)=-i\int d\Sigma^{\mu}(\psi\overleftrightarrow{\partial_{\mu}}\phi^{*})~.\label{eq:kknorm}\end{equation}

\subsection{Flat slicing}

The spacetime metric for $d$-dimensional de Sitter is \begin{equation}
ds^{2}=-dt^{2}+e^{-2t}dx_{d-1}^{2}~,\label{eq:flatmet}\end{equation}
where $dx_{d-1}^{2}$ is the standard metric on flat ($d-1$)-dimensional
Euclidean space. The coordinates only cover the lower triangle of
de Sitter space, the time-reverse of the inflating patch. 

Unlike the case of global coordinates spatial slices are now non-compact,
so it is most convenient to make mode functions plane-wave normalizable.
We therefore choose a basis for the spatial mode functions $e^{ik\cdot x}$
where $k\in\mathbb{R}^{d-1}$ so that\begin{equation}
\phi(x,t)=\int\frac{dk^{d-1}}{(2\pi)^{d-1}}~a_{k}f_{k}(t)e^{ik\cdot x}+a_{k}^{\dagger}f_{k}^{*}(t)e^{-ik\cdot x}~,\label{eq:modflat}\end{equation}
where the time-dependent mode functions are now \cite{Birrell}\[
f_{k}(t)=\frac{\sqrt{\pi}}{2}e^{\frac{\pi\mu}{2}}e^{\frac{d-1}{2}t}H_{i\mu}^{(2)}(|k|e^{t})~,\]
with $H_{i\mu}^{(2)}$ a Hankel function of the second kind. The modes
$f_{k}(t)e^{ik\cdot x}$ have a plane-wave normalization with respect
to the Klein-Gordon inner product (\ref{eq:kknorm})\[
(k,k')=(2\pi)^{d-1}\delta^{d-1}(k-k')~.\]

In the limit $t\to-\infty$\begin{eqnarray}
f_{k}(t) & \sim & A_{k}e^{h_{-}t}+B_{k}e^{h_{+}t}\nonumber \\
A_{k} & = & \frac{i\Gamma(i\mu)}{2\sqrt{\pi}}e^{\frac{\pi\mu}{2}}\left(\frac{|k|}{2}\right)^{-i\mu},\qquad B_{k}=\frac{i\Gamma(-i\mu)}{2\sqrt{\pi}}e^{-\frac{\pi\mu}{2}}\left(\frac{|k|}{2}\right)^{i\mu}~.\label{eq:flatmod}\end{eqnarray}
Note the much simpler $k$ dependence versus (\ref{eq:sphmod}). The
following identity is useful in the manipulation of these mode functions:\[
\Gamma(z)\Gamma(1-z)=\frac{\pi}{\sin\pi z}~.\]

\section{Two dimensions}

\subsection{Asymptotic symmetry group}

We will first work with two-dimensional de Sitter in global coordinates,\[
ds^{2}=-d\tau^{2}+\cosh^{2}\tau~d\theta^{2}~,\]
 with $\theta$ periodically identified up to $2\pi$. Consider the
group of diffeomorphisms that leave fixed the asymptotic conditions\begin{eqnarray}
g_{\tau\tau} & = & -1+\mathit{O}\left(e^{2\tau}\right)\nonumber \\
g_{\tau\theta} & = & \mathit{O}\left(1\right)\nonumber \\
g_{\theta\theta} & = & \frac{1}{2}e^{-2\tau}+\mathit{O}\left(1\right)\label{eq:metriccon}\end{eqnarray}
at past infinity $\tau\to-\infty$. This group is generated by the
vector fields whose asymptotic behavior is\begin{equation}
L_{U}=U(\theta)\frac{\partial}{\partial\theta}+U'(\theta)\frac{\partial}{\partial\tau}+\mathit{O}\left(e^{2\tau}\right),\label{eq:ldiff}\end{equation}
where $U(\theta)$ is an arbitrary function on the circle. For the
choice $U(\theta)=ie^{in\theta}$, the $L_{n}$'s generate a centerless
Virasoro (or Witt) algebra\[
[L_{m},L_{n}]=(m-n)L_{m+n}~.\]
$L_{0},L_{\pm1}$ are the vector fields associated with the $dS_{2}$
isometry group $SL(2,\mathbb{R})$. Note that this choice of global
generators is related by a nontrivial isomorphism to the choice studied
in \cite{Bousso:2001mw,Guijosa:2003ze,Lowe:2004nw}. In comparing
formulas in the two notations, bear in mind that \begin{eqnarray*}
\tilde{L}_{0} & = & \frac{1}{2}\left(L_{-1}-L_{1}\right)\\
\tilde{L}_{1} & = & -\frac{i}{2}\left(L_{1}+L_{-1}+2L_{0}\right)\\
\tilde{L}_{-1} & = & \frac{i}{2}\left(L_{1}+L_{-1}-2L_{0}\right)~,\end{eqnarray*}
where the $\tilde{{L}}_{n}$ are the Virasoro generators of \cite{Bousso:2001mw,Guijosa:2003ze,Lowe:2004nw}.
Therefore the generator of time translations in the static patch will
be $(L_{-1}-L_{1})/2$ in the present notation. Furthermore, in the
present notation, Hermitian conjugation will act as $L_{n}\to L_{-n}$
, as in conventional CFT.

The action of the asymptotic symmetry group on the gravitational degrees
of freedom is generated by a boundary stress energy tensor studied
by Brown and York \cite{Brown:1992br}. A version of this for two-dimensional
de Sitter, realized as a solution of dilaton gravity, has been studied
in \cite{Cadoni:2002kz,Medved2003}. The algebra generated by moments
of this stress tensor is Virasoro, with a nontrivial central charge.

\subsection{Representations of Virasoro algebra}

It is helpful to first briefly review the AdS/CFT case. In the simplest
situation, an operator transforming as a conformal primary with positive
real conformal weight is mapped to a single particle mode in the bulk
of AdS. The isometries of AdS are identified with elements of the
global conformal group on the boundary, and the particle modes form
a representation of this group. For $AdS_{2}$ this isometry group
is enhanced to a larger asymptotic symmetry group which is the full
Virasoro algebra, and $L_{0}$ generates light-cone time translations
\cite{Strominger1999}. Furthermore, descendant operators correspond
to multi-particle states with larger values of $L_{0}$. Thus a highest
weight Virasoro module maps into multiparticle states in the bulk,
which are descendants of the lightest state \cite{Maldacena1998}. 

Let us follow the same strategy in de Sitter. We begin with a scalar
field excitation in the bulk, and fix the scalar mass $m$. We will
then proceed to act on this state with the Virasoro generators, and
deduce the structure of the full Virasoro module. An important difference
with AdS is immediately apparent: $L_{0}=i\frac{\partial}{\partial\theta}$
generates translations around the spacelike circle, so the eigenvalues
of $L_{0}$ will be unbounded above and below.

The scalar field stress energy tensor is\[
T_{\mu\nu}=\partial_{\mu}\phi\partial_{\nu}\phi-\frac{1}{2}g_{\mu\nu}g^{\lambda\rho}\partial_{\rho}\phi\partial_{\lambda}\phi-\frac{1}{2}g_{\mu\nu}m^{2}\phi^{2}~,\]
so the matter contribution to the Virasoro generators (\ref{eq:ldiff})
takes the form

\begin{equation}
L_{U}=-\frac{i}{2}e^{-\tau}\int_{0}^{2\pi}d\theta~\left(U(\theta)\phi'\dot{\phi}+\frac{1}{2}U'\left(\dot{\phi}^{2}+m^{2}\phi^{2}\right)\right)~,\label{eq:vgens}\end{equation}
where we have dropped terms subleading as $\tau\to-\infty$.

To make this expression well-defined, we must specify a normal-ordering
prescription. We define normal ordering with respect to the Bunch-Davies,
or Euclidean, vacuum state. The mode decomposition is (\ref{eq:modede})
-- and we will set $l=|k|$, where $k\in\mathbb{Z}$ is the quantum
number of momentum around the spatial circle. 

We normal order with respect to this vacuum state, i.e., we order
$a_{k}$ to the right of $a_{k}^{\dagger}$'s when defining the generators
(\ref{eq:vgens}), and we denote the normal ordered generators $:L_{U}:$.
It is straightforward to verify the $:L_{U}:$'s generate an algebra
of Virasoro type. The only potential subtlety is computing the central
charge. This is discussed in more detail in appendix A, where it is
shown the central charge vanishes. 

Time-dependent coefficients seem to appear when the generators (\ref{eq:vgens})
are written in terms of creation and annihilation operators, using
the mode expansion (\ref{eq:modede}) and the asymptotic form (\ref{eq:sphmod}).
However this time dependence actually completely cancels out of the
generators, and the resulting coefficients are recorded in appendix
B. The generic $:L_{n}:$ involves terms of the three basic types
$a^{\dagger}a$, $a^{\dagger}a^{\dagger}$ and $aa$, and consequently
mixes states with different particle number (notice that states with
an even/odd number of particles transform irreducibly). For the global
($n=0,\pm1)$ generators, however, the terms of the second and third
type can be seen to cancel out, implying that single-particle states
by themselves carry a representation of the isometry group, in consonance
with the results of \cite{Chernikov:1968zm,Tagirov:1972vv,Leblond:2002tf,Balasubramanian:2002zh,Guijosa:2003ze,Lowe:2004nw}.

One might wonder if the different $\alpha$-vacua of field theory
in de Sitter give rise to different representations. As discussed
in \cite{Guijosa:2003ze} at the level of the global conformal group
$SL(2,\mathbb{R})$, the Euclidean vacuum is picked out by demanding
the bulk-to-boundary mapping be local in the boundary coordinates.
This jibes well with the expected instabilities of interacting $\alpha$-vacua
once coupling to bulk gravity is included \cite{Goldstein:2003qf}.

The bulk-to-boundary map was described in \cite{Guijosa:2003ze} for
the case of single-particle states. There the $dS_{2}/CFT_{1}$ mapping
is performed by noticing there are two equivalent representations
expressed in terms of functions of $\theta,\tau$ decomposed in modes
$f_{k}(\tau)e^{ik\theta}$ or just in terms of {}``boundary'' modes
$e^{ik\theta}$. The factors $f_{k}(\tau)$ can then be viewed as
the bulk-to-boundary propagator in momentum space. In the present
context, time-independence of the conformal generators when written
in terms of creation and annihilation operators $a_{k}$, $a_{k}^{\dagger}$
is important for the consistency of these dual pictures. 

The new feature with the full Virasoro representations is that states
with different particle number are part of the same representation.
Nevertheless, once the generators are written as time-independent
linear combinations quadratic in creation/annihilation operators,
the same basic picture of the bulk-to-boundary mapping will carry
over where $f_{k}(\tau)$ factors play the role of bulk-to-boundary
propagators. We reiterate we are only treating a free scalar field,
so interactions are ignored. Generically we will get distinct representations
corresponding to Virasoro acting on distinct multiparticle excitations
of the Euclidean vacuum as summarized in table \ref{cap:The-de-Sitter/CFT}. 

\begin{table}
\begin{tabular}{|c|c|}
\hline 
Bulk dS&
Boundary CFT\tabularnewline
\hline
\hline 
$\left\{ L_{n}\right\} |E\rangle$&
$1$+descendants\tabularnewline
\hline 
$\left\{ L_{n}\right\} f_{k}^{*}(\tau)a_{k}^{\dagger}|E\rangle$&
$\mathcal{O}_{k}$+descendants\tabularnewline
\hline 
$\left\{ L_{n}\right\} f_{k}^{*}(\tau)f_{k'}^{*}(\tau)a_{k}^{\dagger}a_{k'}^{\dagger}|E\rangle$&
$\mathcal{O}_{k,k'}$+descendants\tabularnewline
\hline
\end{tabular}

\caption{The de Sitter state/CFT operator mapping\label{cap:The-de-Sitter/CFT}.
Here $\left\{ L_{n}\right\} $ denotes an element in the universal
enveloping algebra of Virasoro. The CFT operator $\mathcal{O}_{k}$
denotes a component of a principal series representation of $SL(2,\mathbb{R})$
with spatial momentum $k$. The operator $\mathcal{O}_{k,k'}$ denotes
a component in the tensor product of two principal series representations
of $SL(2,\mathbb{R})$ labelled by spatial momenta $k$ and $k'$.
A similar pattern continues for higher tensor products.}
\end{table}

\subsection{Unitarity}

Ordinarily, one concludes that unitary representations of Virasoro
with vanishing central charge are trivial. However the representation
discussed here is not of highest weight type, so let us re-examine
this conclusion. 

The symmetry algebra changes particle number, so in particular the
vacuum is not invariant. The representation space lives in a subspace
of the Fock space of free particles in the Euclidean vacuum. The natural
norm on this entire Fock space is defined in momentum space as\begin{equation}
\prod_{k=-\infty}^{\infty}\left\langle n'_{k}|n_{k}\right\rangle =\prod_{k=-\infty}^{\infty}\delta_{n_{k},n'_{k}}~,\label{eq:mnorm}\end{equation}
where $n_{k}$ is the occupation number of mode $k$ and\[
|n_{k}\rangle=\frac{1}{\sqrt{n_{k}!}}a_{k}^{\dagger n_{k}}|E\rangle\]
with $|E\rangle$ the Euclidean vacuum. The notion of Hermitian conjugation
induced by this norm on the Virasoro generators is $L_{n}^{\dagger}=L_{-n}$.
The states $L_{n}|\psi\rangle$ therefore have positive norm, for
$|\psi\rangle$ a state with finite particle number. In particular\[
\langle E|L_{-n}L_{n}|E\rangle=2n\langle E|L_{0}|E\rangle+\langle E|L_{n}L_{-n}|E\rangle~,\]
and while the first term on the right hand side vanishes, the second
term is non-vanishing. For a highest weight representation, on the
other hand, both terms would vanish. In that case, the representation
that includes the vacuum is nontrivial only for nonzero central charge.

\subsection{Planar coordinates}

The above discussion of Virasoro generators can be generalized to
planar coordinates\[
ds^{2}=-dt^{2}+e^{-2t}dz^{2}\]
where $z\in\mathbb{R}$. The main difference is finding the most convenient
choice of mode functions, and choosing a basis for the Virasoro generators.
Unlike the case of global coordinates spatial slices are now non-compact,
so it is most convenient to make mode functions plane-wave normalizable.
We therefore choose a basis for the spatial mode functions $e^{ikz}$
where $k\in\mathbb{R}$ as in (\ref{eq:modflat}).

To keep the action of the Virasoro generators on these mode functions
as simple as possible, we likewise choose a basis $U(z)=ie^{ipz}$
where $p\in\mathbb{R}$. The main new feature is that Virasoro generators
are labelled by a real parameter $p$ rather than the familiar integer
parameter. Otherwise the calculations of the previous subsections
go through with sums replaced by integrals and only minor modifications.
Explicit expressions for the Virasoro generators are given in appendix
B. Note that if we instead choose the usual basis of Virasoro generators
$U=-z^{n}$ the action of the generators will take us outside the
space of plane-wave normalizable functions.

\section{Three dimensions}

\subsection{Asymptotic symmetry group}

For the case of three-dimensional de Sitter, it is more convenient
to use flat spatial sections, to exhibit the left-right decomposition
of the $CFT_{2}$. Introducing $z=x_{1}+ix_{2}$ versus (\ref{eq:flatmet}),
the metric is \[
ds^{2}=-dt^{2}+e^{-2t}dz\, d\bar{z}~.\]
 We consider the asymptotic symmetry group leaving fixed the conditions
\cite{Strominger:2001pn}\begin{eqnarray*}
g_{z\bar{z}} & = & \frac{1}{2}e^{-2t}+O(1)\\
g_{tt} & = & -1+O(e^{2t})\\
g_{zz} & = & O(1)\\
g_{tz} & = & O(1)\end{eqnarray*}
as $t\to-\infty$. This symmetry group is the group of diffeomorphisms
generated by the vector fields\[
\zeta_{U}=U\partial_{z}+\frac{1}{2}\partial_{z}U\partial_{t}+O(e^{2t})+\mathrm{complex}~\mathrm{conjugate}~,\]
where $U$ is holomorphic. The basis choice $U(z)=-z^{n+1}$ yields
the standard mutually commuting sets of Virasoro generators $L_{n},\bar{L}_{n}$.

\subsection{Representations of Virasoro left+right?}

By analogy with the case of $dS_{2}$ analyzed in the previous section,
we might expect the multiparticle states associated with a bulk scalar
field to belong to a representation of left+right Virasoro that decomposes
into principal series representations of the global isometry group
$SL(2,\mathbb{C})$. The Virasoro generators now take the form\begin{equation}
L_{U}=-i~e^{-2t}\int d^{2}z\,\left(U\partial_{z}\phi\partial_{t}\phi+\frac{1}{4}\partial_{z}U\left(\left(\partial_{t}\phi\right)^{2}+m^{2}\phi^{2}\right)\right)\label{eq:vgens3d}\end{equation}
(and likewise for $\bar{L}_{U}$), again dropping terms that vanish
as $t\to-\infty$. The analog of the calculation described in appendix
A implies these operators generate a Virasoro algebra with vanishing
central charge. These operators have a well-defined action on local
operators in the CFT.

However, since the function $U(z)$ necessarily has non-simple poles
for all but the global $SL(2,\mathbb{C})$ transformations, the spatial
integral will diverge for states with smooth distributions of matter,
such as we have for a one-particle excitation of the Euclidean vacuum.
(The divergence is of course also visible in momentum space, if we
attempt to expand the generators in terms of creation and annihilation
operators as in appendix B.) So, even though the $L_{n}$ operators
(\ref{eq:vgens3d}) satisfy the expected commutation relations, their
matrix elements between generic multiparticle states are ill-defined
unless $n=0,\pm1$. 

The key issue here is the choice of boundary conditions. Recall that
the Brown-York tensor $\tau_{ij}$ (with $i,j$ running over the spatial
directions) \cite{Brown:1992br} satisfies\[
D_{i}\tau^{ij}=-T^{0j}~,\]
where $T^{\mu\nu}$ is the matter stress energy tensor. This equation
may be integrated across a two-dimensional region at past infinity
to give\begin{equation}
H_{C_{1}}(\zeta)-H_{C_{2}}(\zeta)=-e^{-2t}\int_{\Sigma}d^{2}z~T^{0i}\zeta_{j}~,\label{eq:bycharge}\end{equation}
with\[
H_{C}(\zeta)=\frac{1}{2\pi i}\oint_{C}dz~\tau_{zz}\zeta^{z}~.\]
If we assume matter stress energy is negligible at past infinity,
the right hand side of (\ref{eq:bycharge}) vanishes, and the $H$'s
are contour-independent %
\footnote{If the dual CFT were to be radially quantized, as is standard, e.g.,
in string theory applications, this contour-independence would of
course identify the $H$'s as conserved charges. We would like to
emphasize, however, that in the $dS_{d}/CFT_{d-1}$ context there
is a priori no time direction in the CFT, and so spatial integrals
are naturally $(d-1)$- rather than $(d-2)$-dimensional. %
}. They generate an algebra that is a direct sum of holomorphic and
anti-holomorphic copies of the Virasoro algebra with a non-trivial
central charge \cite{Strominger:2001pn,Klemm:2001ea}. 

However, if we want to describe even a one-particle state, then the
matter stress-energy is not completely negligible at past infinity,
and so we cannot drop the terms on the right-hand side of (\ref{eq:bycharge}).
As we have already noted, for all but the global $SL(2,\mathbb{C})$
transformations the integral over $\Sigma$ will be ill-defined. We
conclude therefore that if we allow for such boundary conditions on
the matter fields, then the asymptotic symmetry group collapses to
$SL(2,\mathbb{C})$ and there is no sensible extension of the principal
series representations to the algebra of left/right Virasoro.

Note that the crucial difference with the two-dimensional case analyzed
in the previous section is that the $dS_{2}$ asymptotic symmetry
group is in fact the full group of diffeomorphisms on the circle,
for which one can choose a normalizable basis and obtain well-defined
integrals.

\section{Discussion}

Since we have described the mapping from representations of Virasoro
to particle states in de Sitter, as summarized in table \ref{cap:The-de-Sitter/CFT},
we can identify in the conformal field theory the procedure leading
to entropy counting. States in the asymptotic past will form representations
of the isometry group and can be propagated to arbitrary times using
the isometry generators. 

For definiteness, let us restrict attention to the case of $dS_{2}$
in spherical coordinates, though generalization to the flat slicing
is straightforward. One may choose a basis of these states which diagonalizes
the generator of time translations in the static patch, $(L_{-1}-L_{1})/2$.
This change of basis is accomplished as follows. Positive frequency
modes with respect to global time $\phi^{E}$ are analytic in the
lower half plane when expressed in terms of Kruskal coordinates. These
modes are decomposed into modes with positive and negative frequencies
with respect to time in the south and north static patches \cite{Unruh:1976db}.
The decomposition takes the following form \cite{Bousso:2001mw}\begin{equation}
\phi_{\omega}^{E}=\phi_{\omega}^{S}+e^{-\pi\omega}\phi_{\omega}^{N}~,\label{eq:eucmoda}\end{equation}
i.e., the positive-frequency component $\phi_{\omega}^{S}$ comes
from the southern causal diamond, while the negative-frequency component
$\phi_{\omega}^{N}$ comes from the northern causal diamond, where
$t$ runs backward. There is a second linear combination that has
positive frequency with respect to global time,\begin{equation}
\tilde{\phi}_{\omega}^{E}=\phi_{\omega}^{N*}+e^{-\pi\omega}\phi_{\omega}^{S*}~,\label{eq:eucmodb}\end{equation}
where the roles of north and south are swapped.

Now both $\phi_{\omega}^{E}$ and $\tilde{\phi}_{\omega}^{E}$ annihilate
the Euclidean vacuum when $\omega>0$. It is then possible to show
the Euclidean vacuum can be written as\begin{eqnarray*}
|E\rangle & = & \prod_{\omega>0}(1-e^{-2\pi\omega})^{1/2}\exp\left(e^{-\pi\omega}a_{\omega}^{S\dagger}a_{\omega}^{N}\right)|S\rangle\otimes|N\rangle\\
 & = & \prod_{\omega>0}(1-e^{-2\pi\omega})^{1/2}\sum_{n_{\omega}=0}^{\infty}e^{-\pi\omega n_{\omega}}|n_{\omega},S\rangle\otimes|n_{\omega},N\rangle\end{eqnarray*}
where $|N\rangle$ is the northern vacuum annihilated by $\phi_{\omega}^{N*}$
and $|S\rangle$ is the southern vacuum annihilated by $\phi_{\omega}^{S}$.
The state $|n_{\omega},S\rangle$ denotes occupation number $n_{\omega}$
in the $\omega$ mode with respect to the southern vacuum state, and
likewise for $|n_{\omega},N\rangle$. It is then straightforward to
trace over modes in the northern diamond, to obtain a density matrix
for modes within only the southern causal diamond\begin{equation}
\rho_{S}=\prod_{\omega>0}(1-e^{-2\pi\omega})\sum_{n_{\omega}=0}^{\infty}e^{-2\pi\omega n_{\omega}}|n_{\omega},S\rangle\otimes\langle n_{\omega},S|~.\label{eq:densmat}\end{equation}
This takes the form of a thermal density matrix, where the units are
such that the temperature of de Sitter is $1/2\pi$. The free energy
associated with this entanglement is \begin{equation}
e^{-2\pi F}=\mathrm{Tr}~e^{-2\pi\omega n_{\omega}}~.\label{eq:freeen}\end{equation}

Using the unitary Virasoro representations we have constructed, we
now can map each step in the above entropy calculation into a counting
of operators in the boundary CFT with a particular weighting. For
example, the conjugate of each of the modes (\ref{eq:eucmoda}) and
(\ref{eq:eucmodb}) is dual to a CFT operator $\mathcal{O}_{\omega}^{E}$
or $\tilde{\mathcal{O}}_{\omega}^{E}$ which is a linear combination
of the operators dual to single particle modes described above. These
operators create modes with correlations between the north and south
static patches. Note that since the bulk theory is formulated around
the Euclidean vacuum, it is not possible to avoid this correlation
and create modes localized purely in one static patch. A similar mapping
will exist for all the multiparticle states. From the CFT viewpoint,
the partition function (\ref{eq:freeen}) corresponds to the following
weighted sum over the operator content of the CFT\[
e^{-2\pi F}=\mathrm{Tr}~e^{-\pi i(L_{-1}-L_{1})}~,\]
where the trace is over the full set of operators dual to the scalar
field Fock space, as described in table \ref{cap:The-de-Sitter/CFT}.
The trace is defined with the restriction that the eigenvalue of $i(L_{-1}-L_{1})$
is positive, corresponding to the $\omega>0$ condition in (\ref{eq:densmat}).
This at least solves one aspect of the entropy counting problem in
dS/CFT.

The problem is that the resulting entropy is divergent, due to a continuum
of states with finite $i(L_{-1}-L_{1})$ localized near the cosmological
horizon. These may be cutoff with analogs of the brick-wall cutoff
of 't Hooft \cite{'tHooft:1985re}. Such calculations have been performed
in \cite{Kim:1998zs}. All these may now be interpreted in terms of
the boundary CFT using the dictionary we have provided -- but we need
new physical input to really understand the role of the cutoff. We
hope to explore these issues in the future, applying ideas of $q$-deformation
\cite{Guijosa:2003ze,Lowe:2004nw,Pouliot:2003vt,Banks:2005bm}.

\begin{acknowledgments}
We thank Y.A. Neretin for helpful correspondence. The work of A.G.
is supported by Mexico's National Council for Science and Technology
(CONACyT) grants 40754-F and J200.315/2004 and by DGAPA-UNAM grant
IN104503-3. The research of D.A.L. is supported in part by DOE grant
DE-FE0291ER40688-Task A and NSF U.S.-Mexico Cooperative Research grant
\#0334379. J.M is supported by an overseas postdoctoral fellowship
of the NRF (South Africa) and thanks Kenneth Hughes, Simon Judes and
Amanda Weltman for useful discussions.
\end{acknowledgments}
\appendix

\section{Central charge}

For $U=ie^{in\theta}$, we obtain\begin{equation}
[:L_{U}:,:L_{U^{*}}:]=\cosh t\int d\theta~d\theta'~e^{in(\theta'-\theta)}i\delta'(\theta-\theta')\left(\dot{\phi}(\theta)\phi'(\theta')+\dot{\phi}(\theta')\phi'(\theta)\right)~.\label{eq:vircom}\end{equation}
To make the normal ordering terms on the right hand side of (\ref{eq:vircom})
well-defined, one may use a $i\epsilon$ prescription. It is also
necessary to regulate the $\delta'(\theta-\theta')$ factor using
the same prescription. A potential central term in the algebra then
arises from $\left\langle \dot{\phi}(\theta)\phi'(\theta')+\dot{\phi}(\theta')\phi'(\theta)\right\rangle $
in the limit that points coincide. In this limit, we can evaluate
using\[
\left\langle \phi(\theta,t)\phi(\theta',t')\right\rangle =-\frac{1}{4\pi}\log((t-t'-i\epsilon)^{2}-\cosh t\cosh t'(\theta-\theta')^{2})~.\]
 Direct computation for the correlator of interest shows at equal
times the $i\epsilon$ terms cancel and\[
\left\langle \dot{\phi}(\theta)\phi'(\theta')+\dot{\phi}(\theta')\phi'(\theta)\right\rangle =0~.\]
 Therefore the central term vanishes.

\section{Virasoro generators}

In this appendix we compute the expansion of the Virasoro generators
in terms of creation and annihilation operators, and demonstrate the
time-independence of the coefficients. Consider first the case of
$dS_{2}$ in spherical coordinates, where the mode expansion is given
by (\ref{eq:modede}) with $l\to k\in\mathbb{Z}$ and $Y_{k}(\theta)=e^{ik\theta}/\sqrt{2\pi}$
. Using the asymptotic form (\ref{eq:sphmod}) and substituting into
(\ref{eq:vgens}) with $U=ie^{in\theta}$, the Virasoro generator
$:L_{n}:$ is expressed as a sum over $k$ of an expression quadratic
in creation/annihilation operators (which by definition is meant to
be normal-ordered). At first sight the coefficients in this sum appear
to have some remaining time-dependence, but this dependence cancels
when one combines the two distinct terms in the sum that contribute
to a given $aa$, $a^{\dagger}a$, or $a^{\dagger}a^{\dagger}$ product,
and makes use of the identity $h_{\pm}-h_{\pm}^{2}-m^{2}=0$. The
end result is \begin{eqnarray*}
2i:L_{n}: & = & \sum_{k\ge[(n+1)/2]}a_{k}^{\dagger}a_{-k+n}^{\dagger}\left[\left(n(h_{-}-2m^{2})+2i\mu k\right)A_{-k+n}^{*}B_{k}^{*}+\left(n(h_{+}-2m^{2})-2i\mu k\right)A_{k}^{*}B_{-k+n}^{*}\right]\\
 &  & +\sum_{k\ge[(n+1)/2]}a_{-k}a_{k-n}\left[\left(n(h_{+}-2m^{2})-2i\mu k\right)A_{k-n}B_{-k}+\left(n(h_{-}-2m^{2})+2i\mu k\right)A_{-k}B_{k-n}\right]\\
 &  & +\sum_{k=-\infty}^{\infty}a_{n-k}^{\dagger}a_{-k}\left[\left(n(h_{-}-2m^{2})+2i\mu k\right)A_{n-k}^{*}A_{-k}+\left(n(h_{+}-2m^{2})-2i\mu k\right)B_{n-k}^{*}B_{-k}\right]\end{eqnarray*}
If $n$ is even, the first term of the sums in the first and second
lines corresponds to $k=n/2$, and should actually have a coefficient
that is only half as big as the other terms. 

For $dS_{2}$ in planar coordinates (where the modes are given by
(\ref{eq:modflat}) and (\ref{eq:flatmod})) one finds a completely
analogous result, with the integer $n$ that labels the Virasoro generator
replaced by a real number $p$, the sums over $k$ replaced by integrals
$\int_{(p+1)/2}^{\infty}dk/2\pi$ and $\int_{-\infty}^{\infty}dk/2\pi$,
and the factor of $2$ removed from the left-hand side of the equation.

It should be noted the coefficients of the $a^{\dagger}a^{\dagger}$
and the $aa$ terms vanish for the global generators $L_{0,\pm1}$
for $dS_{2}$ in spherical coordinates. The same is true for the global
generators in flat coordinates, but these are nontrivial linear combinations
in the basis of Virasoro generators defined above, corresponding to
$U=-1,-z,-z^{2}$. The global generators therefore preserve particle
number, and one sees the one-particle sector of the Fock space transforms
as a unitary principal series representation, as expected.

\bibliographystyle{brownphys}
\clearpage\addcontentsline{toc}{chapter}{\bibname}\bibliography{qdscft}

\end{document}